\documentclass[11pt]{article}
\usepackage{times}
\usepackage{fullpage}
\usepackage{amsmath}
\usepackage{amssymb}
\usepackage{amsfonts}
\usepackage[all,poly]{xy}

\newcounter{algoctr}

\newif\ifnotesw\noteswtrue
   {\ifnotesw\marginpar[\hfill\(\top\)]{\(\top\)}\fi}%
   {\ifnotesw\marginpar[\hfill\(\bot\)]{\(\bot\)}\fi}

\newcommand{\mnote}[1]%
    {\ifnotesw\marginpar%
        [{\scriptsize\begin{minipage}[t]{\marginparwidth}
        \raggedleft#1%
                        \end{minipage}}]%
        {\scriptsize\begin{minipage}[t]{\marginparwidth}
        \raggedright#1%
                        \end{minipage}}%
    \fi}

\newcommand{\ignore}[1]{}

\newcommand{\etal}{{\it et al. }}

\newsavebox{\given}
\savebox{\given}[1em]{\rule[-1.5ex]{.2mm}{4ex}}

\newcommand{\bnum}{\begin{equation}}
\newcommand{\enum}{\end{equation}}

\newtheorem{theorem}{Theorem}
\newtheorem{corollary}[theorem]{Corollary}
\newtheorem{lemma}[theorem]{Lemma}

\newtheorem{conjecture}{Conjecture}

\newtheorem{definition}{Definition}

\newcommand{\blackslug}{\rule{7pt}{7pt}}

\newcommand{\iverson}[1]{\lbrack\!\lbrack #1 \rbrack\!\rbrack}

\newcommand{\real}{\ifmmode {\rm R} \else ${\rm R}$ \fi}
\newcommand{\nat}{\ifmmode {\rm N} \else ${\rm N}$  \fi}
\newcommand{\tot}{\ifmmode {\cal T} \else ${\cal T}$ \fi}
\newcommand{\sigstar}{\ifmmode \Sigma^{\ast} \else $\Sigma^{\ast}$ \fi}
\renewcommand{\star}{\ast}

\newcommand{\inn}{\ifmmode \in \else $\in$ \fi}
\renewcommand{\phi}{\ifmmode \varphi \else $\varphi$ \fi}
\renewcommand{\le}{\ifmmode \leq \else $\leq$ \fi}
\renewcommand{\ge}{\ifmmode \geq \else $\geq$ \fi}
\renewcommand{\ne}{\ifmmode \neq \else $\neq$ \fi}
\newcommand{\lt}{\ifmmode < \else $<$ \fi}
\newcommand{\gt}{\ifmmode > \else $>$ \fi}
\newcommand{\eq}{\ifmmode = \else $=$ \fi}
\newcommand{\half}{\ifmmode \frac{1}{2} \else $\frac{1}{2}$ \fi}
\newcommand{\oneovern}{\ifmmode \frac{1}{n} \else $\frac{1}{n}$ \fi}

\newcommand{\ra}{\ifmmode \rightarrow \else $\rightarrow$ \fi}
\newcommand{\qed}{\hfill{\setlength{\fboxsep}{0pt}
\framebox[7pt]{\rule{0pt}{7pt}}}}

\renewcommand{\notin}{\ifmmode \not\in \else $\not\in$ \fi}
\newlength{\thislabel}
\newcommand{\labsize}[1]{\settowidth{\thislabel}{#1}}

\newcommand{\prf}{\par\noindent{\sl Proof } \hspace{.01 in}}
\newcommand{\zo}{\{0,1\}}

\newcommand{\ep}{\mbox{\bf E}}

\newcommand{\lip}{\langle}
\newcommand{\rip}{\rangle}
\def\Nat{\mathbb N}
\def\Complex{\mathbb C}
\def\Int{\mathbb Z}
\def\Real{\mathbb R}

\newcommand{\bra}[1]{\lip #1 |}
\newcommand{\ket}[1]{| #1 \rip}
\newcommand{\braket}[2]{\lip #1 | #2 \rip}
\newcommand{\tderiv}[1]{\frac{\mathsf{d}}{\mathsf{d}t} #1}
\newcommand{\dt}{\mathsf{d}t}

\title{
{\bf A note on graphs resistant to quantum uniform mixing}
} 
\author{
{William Adamczak}		
\and {Kevin Andrew}		
\and {Peter Hernberg}		
\and {Christino Tamon}\footnote{Contact author: tino@clarkson.edu}		
}
\date{}

\begin{document}
\bibliographystyle{plain}
\maketitle

\begin{abstract}
Continuous-time quantum walks on graphs is a generalization of continuous-time Markov chains on discrete structures.
Moore and Russell proved that the continuous-time quantum walk on the $n$-cube is instantaneous exactly uniform 
mixing but has no average mixing property. 
On complete (circulant) graphs $K_{n}$, the continuous-time quantum walk 
is neither instantaneous (except for $n=2,3,4$) nor average uniform mixing (except for $n=2$). 
We explore two natural {\em group-theoretic} generalizations of the $n$-cube as a $G$-circulant 
and as a bunkbed $G \rtimes \Int_{2}$, where $G$ is a finite group.
Analyses of these classes suggest that the $n$-cube might be special in having instantaneous uniform mixing and
that non-uniform average mixing is pervasive, i.e., no memoryless property for the average limiting distribution;
an implication of these graphs having zero spectral gap.
But on the bunkbeds, we note a memoryless property with respect to the two partitions.
We also analyze average mixing on complete paths, where the spectral gaps are nonzero.
\end{abstract}

\section{Introduction}

In quantum information processing, a quantum analogue of classical random walks has been the focus of study in a sequence 
of works \cite{fg,cfg,abnvw,aakv,mr,k}. There are two interesting variants of this quantum analogue of random walks on graphs. 
The first variant is the discrete quantum walk model studied originally by Ambainis \etal \cite{abnvw} and Aharonov \etal \cite{aakv}.
Many similarities have been shown between classical and quantum walks in this discrete setting.
On the other hand, a quantum variant of continuous-time Markov chains, introduced by Farhi and Gutmann \cite{fg}, has shown 
more differences in classical versus quantum behavior. This was exploited beautifully in a recent work by Childs \etal \cite{ccdfgs}
where an exponential computational speedup was achieved via continuous-time quantum walks. 

An earlier work by Moore and Russell \cite{mr} studied continuous-time quantum walk on the $n$-cube. They proved that there are times
when the probability distribution corresponding to the quantum walk is exactly uniform. 
This property is called {\em instantaneous exact uniform mixing}. In \cite{abtw}, it is shown that this property is not shared
by some circulant graphs, notably the large complete graphs $K_{n}$ (except for $K_{2}$, $K_{3}$, and $K_{4}$).
Recently, Gerhardt and Watrous \cite{gw} proved that the non-uniform mixing also afflicts the important Cayley graphs 
of the symmetric groups $S_{n}$ (except for the trivial case of $S_{2}$). 

The focus of this note is on continuous-time quantum walk on graphs that are generalizations of the circulants and the $n$-cube. 
Our goal is to understand better the continuous-time quantum mixing behavior on these highly symmetric graphs. The two natural
group-theoretic generalizations are as follows.
Given a group $G$, a matrix $M_{G}$ is called a $G$-circulant \cite{diaconis} if $M_{G}[g,h] = f(gh^{-1})$, for some
function $f: G \rightarrow \Complex$. When $G = \Int_{n}$, we get the usual circulant graphs, and when $G = \Int_{2}^{n}$,
we get the $n$-cube. The characters of $G$ are helpful in analyzing the spectra of $M_{G}$ whenever $f$ is a class function,
i.e., a function constant on the conjugacy classes of $G$.

The second generalization arises from groups that are semidirect products $G \rtimes \Int_{2}$, where $G$ is an arbitrary group.
We recover the $n$-cube by letting $G = \Int_{2}^{n-1}$. Also, certain Cayley graphs of the dihedral group 
$D_{n} = \Int_{n} \rtimes \Int_{2}$ and the symmetric group $S_{n} = A_{n} \rtimes \Int_{2}$ fall under this category. 
We use the term {\em bunkbed} to describe a Cayley graph consisting of
two isomorphic copies of a graph connected by parallel edges. Not all interesting Cayley graphs can be viewed in this way;
the natural Cayley graph of the symmetric group under a generating set that are transpositions does not fall under this example.
In another setting, an interesting connection between semidirect products on groups and expansion properties of graphs was
studied by Alon \etal \cite{alw}. The term {\em bunkbed} was coined by H\"{a}ggstrom
in connection with stochastic (physical) processes on product graphs \cite{haggstrom}.

Not many graphs are known to have average uniform mixing. 
An argument given in \cite{aakv} showed that a graph with distinct eigenvalues has average 
uniform mixing under mild assumptions on its eigenvectors. In particular, if the adjacency matrix of the graph is diagonalized 
by either the Fourier or Hadamard matrices, then average uniform mixing holds. The graphs diagonalized by the Fourier 
and Hadamard matrices are the circulant and the hypercubic graphs, respectively.
Some elementary observations show that any Abelian circulants, which include the aforementioned graphs, 
do not have distinct eigenvalues, except for the smallest one, i.e., $K_{2}$.
In fact, the analysis on the average probability of the starting vertex suggests that the continuous-time quantum walk exhibit a
{\em memory} of its initial conditions. More specifically, the average probability of the starting vertex is larger than most
of the other vertices.
Subsequently, we analyze the total variation distance from uniformity of the average distributions arising from 
continuous-time quantum walks on some circulants.

In the analysis of bunkbed graphs, we note a {\em memoryless} property with respect to the 
two isomorphic copies in the bunkbed (a trait inherited seemingly from $K_{2}$).
In the special case of the dihedral group $D_{n}$, the bunkbed is obtained as a Cayley graph using a minimal generating set;
other alternatives for the Cayley graphs include the cycle, the complete bipartite graph 
(obtained using a conjugacy generating set) and the trivial complete graph. 
None of these admit average uniform mixing by a reduction to circulants or a simple analysis on the bunkbed. 

Finally, we consider a natural class of graphs with distinct eigenvalues, namely the complete paths.
Although these graphs distinct eigenvalues, they exhibit non-classical average mixing, i.e., the average limiting
probability of the continuous-time quantum walk does not equal the classical stationary distribution (which is not the uniform
distribution). In a role reversal from the circulant phenomena, the average probability of the starting vertex is
smaller than what is classically expected.

\section{Continuous-time quantum walks}

\ignore{
For $n \in \Nat$, denote $[n] = \{0,1,\ldots,n-1\}$. If $\Psi$ is a logical statement,
then $\iverson{\Psi}$ is $1$ if $\Psi$ is true, and $0$ otherwise.
}
A model of continuous-time quantum walk on graphs was introduced by Farhi and Gutmann \cite{fg} 
(see also \cite{cfg,mr}). 
Let $G = (V,E)$ be a simple, undirected, connected $n$-vertex graph;
we focus only on graphs with these properties.
Let $H$ be the adjacency matrix of $G$ defined as $H[j,k] = \iverson{(j,k) \in E}$, for $j,k \in [n]$,
where $\iverson{S}$ is $1$ if the statement $S$ is true, and $0$ if it is false.
The amplitude wave function $\ket{\psi(t)}$ at time $t$ obeys the Schr\"{o}dinger's equation
\begin{equation}
	i\hslash\tderiv{\ket{\psi(t)}} = H\ket{\psi(t)}
\end{equation}
or (assuming from now on $\hslash = 1$)
$\ket{\psi(t)} = e^{-iHt}\ket{\psi(0)}$.
Assuming the particle starts at vertex $0$, the initial wave function is $\ket{\psi(0)} = \ket{0}$.
The probability that the particle is at vertex $j$ at time $t$ is given by 
\begin{equation} \label{eqn:collapse}
	P_t(j) = |\braket{j}{\psi(t)}|^2.
\end{equation}
The {\em average} probability that the particle is at vertex $j$ is given by
\begin{equation} \label{eqn:average}
	\overline{P}(j) = \lim_{T \rightarrow \infty} \frac{1}{T} \int_{0}^{T} P_{t}(j) \dt.
\end{equation}

Let $\mathsf{Spec}(H)$ be the spectrum of $H$, 
i.e., the set of all eigenvalues $\lambda_{0} \ge \ldots \ge \lambda_{n-1}$ of $H$.
The {\em spectral gap} is defined as $\tau(H) = \min_{j \neq k} |\lambda_{j} - \lambda_{k}|$;
this is non-zero if and only if all eigenvalues are distinct.

Since $H$ is Hermitian, $U_t = e^{-iHt}$ is unitary. Moreover $H$ and $U_{t}$ share
the same orthonormal eigenvectors. If $\lip\lambda_j,\ket{z_j}\rip_{j \in [n]}$ are the eigenvalue and 
eigenvector pairs of $H$, then $\lip e^{-i\lambda_{j}t},\ket{z_j}\rip_{j \in [n]}$ are the eigenvalue and 
eigenvector pairs of $U_t$. So, if $\ket{\psi(0)} = \sum_{j \in [n]} \braket{z_j}{0} \ket{z_j}$, then
\begin{equation} \label{eqn:qwalk}
	\ket{\psi(t)} 
	= e^{-iHt}\sum_{j=0}^{n-1} \braket{z_{j}}{0} \ket{z_{j}}
	= \sum_{j=0}^{n-1} \braket{z_j}{0} e^{-i\lambda_{j}t}\ket{z_j}.
\end{equation}

\par\noindent 
Given two probability distributions $P,Q$ on a finite set $S$,
the {\em total variation distance} between $P$ and $Q$ is defined as
$||P - Q|| = \sum_{s \in S} |P(s) - Q(s)|$.
We now define the two relevant notions of mixing in continuous-time quantum walks on graphs. 

\begin{definition} (instantaneous and average mixing in continuous-time quantum walks \cite{mr,aakv}) \\
Let $G=(V,E)$ be a graph and let $U$ be the uniform distribution on the vertices of $G$.
Let $P_{t}$ and $\overline{P}$ be the instantaneous and average probability distributions of
a continuous-time quantum walk on $G$. For $\epsilon \ge 0$,
\begin{itemize}
\item
$G$ has {\em instantaneous} $\epsilon$-uniform {\em mixing} if there exists $t \in \Real^{+}$ when $||P_{t} - U|| \le \epsilon$.
\item
$G$ has {\em average $\epsilon$-uniform mixing} if 
$||\overline{P} - U|| \le \epsilon$.
\end{itemize}
$G$ has {\em average classical mixing} if the average probability 
$\overline{P}$ equals the stationary distribution of a classical {\em discrete} lazy random
walk on $G$, i.e., the walk $\frac{1}{2}I + \frac{1}{2}A$, where $A$ is the transition probability matrix of $G$.
\end{definition}

\ignore{
A random walk on $G$ is called {\em simple} if the transition probability matrix is 
$H = \frac{1}{d}A$, where $A$ is the adjacency matrix of the $d$-regular graph $G$. The walk
is called {\em lazy} if, at each step, the walk stays at the current vertex with probability 
$\frac{1}{2}$ and moves according to $H$ with probability $\frac{1}{2}$. 
For continuous-time quantum walks, these two notions are {\em equivalent} modulo a time scaling.
The argument is similar to Equation \ref{eqn:phase-factor} by exploiting the commutativity of
$A$ and $I$ that introduces an irrelevant phase factor in the amplitude expression.
}


Next, we state a result from \cite{aakv} showing that a graph with distinct eigenvalues is 
{\em likely} to exhibit average uniform mixing. Their result was stated for the discrete quantum walk
model, but is easily adapted to the continuous-time model.
\ignore{
From Equation (\ref{eqn:qwalk}),
\begin{equation}
\ket{\psi(t)} = e^{-iHt}\sum_{j=0}^{n-1} \braket{z_{j}}{0} \ket{z_{j}}. 
\end{equation}
}

\begin{lemma} \label{lemma:aakv} (Aharonov, Ambainis, Kempe, Vazirani \cite{aakv}) \\
Let $G=(V,E)$ be a simple, $n$-vertex connected, undirected graph with adjacency matrix $H$,
whose eigenvalues are $\lambda_{0} \ge \ldots \ge \lambda_{n-1}$ with corresponding orthonormal 
eigenvectors $\ket{z_{0}}, \ldots, \ket{z_{n-1}}$.
In a continuous-time quantum walk on $G$ starting at vertex $0$, 
the average probability of vertex $\ell$ is
\begin{equation} \label{eqn:aakv}
\overline{P}(\ell) = \sum_{j,k=0}^{n-1} 
		\braket{z_{j}}{0}\braket{0}{z_{k}}\braket{\ell}{z_{j}}\braket{z_{k}}{\ell}
		\iverson{\lambda_{j} = \lambda_{k}}.
\end{equation}
Moreover, if all eigenvalues are distinct, then
$\overline{P}(\ell) = \sum_{j=0}^{n-1} |\braket{\ell}{z_{j}}|^{2} |\braket{z_{j}}{0}|^{2}$.
\end{lemma}
\prf 
The average probability of observing the particle at vertex $\ell$ is
\begin{equation}
\overline{P}(\ell) 
	= 
	\lim_{T \rightarrow \infty} \frac{1}{T} \int_{0}^{T} |\bra{\ell} e^{-itH} \ket{\psi(0)}|^{2} \dt 
	= 
	\sum_{j,k=0}^{n-1} \braket{z_{j}}{0}\braket{0}{z_{k}}\braket{\ell}{z_{j}}\braket{z_{k}}{\ell}
		\lim_{T \rightarrow \infty} \frac{1}{T} \int_{0}^{T} e^{-it(\lambda_{j}-\lambda_{k})} \dt.
\end{equation}
Since $\lim_{T \rightarrow \infty} \frac{1}{T}\int_{0}^{T} e^{-it \Delta} \dt = \iverson{\Delta = 0}$,
this yields the claim.
\ignore{
Thus, we have
\begin{equation} \label{eqn:aakv}
\overline{P}(\ell) = \sum_{j,k=0}^{n-1} 
		\braket{z_{j}}{0}\braket{0}{z_{k}}\braket{\ell}{z_{j}}\braket{z_{k}}{\ell}
		\iverson{\lambda_{j} = \lambda_{k}}.
\end{equation}
}
\qed\\

So, if all eigenvalues are distinct, then average uniform mixing is achieved, under mild conditions on the eigenvectors. 
In particular, there is hope if the graph is diagonalized by the Fourier or Hadamard matrix;
however, these graphs do not have distinct eigenvalues, as we observe later.
Nevertheless, Aharonov \etal \cite{aakv} proved that {\em discrete} quantum walks on odd cycles are average uniform mixing.

\section{Group-theoretic circulants}

Diaconis \cite{diaconis} described a beautiful group-theoretic generalization of circulants. 
Let $G$ be a group of order $n$ and let $f: G \rightarrow \Complex$ be a function defined over $G$.
Consider the matrix $M_{G}^{f}$ defined on $G \times G$ as $M_{G}^{f}[s,t] = f(st^{-1})$.
Modulo the choice of $f$, $G = \Int_{n}$ yields the standard circulants 
and $G = \Int_{2} \times \ldots \times \Int_{2}$ yields the hypercubic graphs
(which includes the $n$-cube as a special case).

Let $\rho: G \rightarrow GL(n,\Complex)$ be a 
representation\footnote{See Serre \cite{serre} for an excellent source on group representation theory.} 
of $G$ with dimension $n$.
The Fourier transform of $f$ at 
$\rho$ is 
\begin{equation}
\hat{f}(\rho) = \sum_{x \in G} f(x)\rho(x).
\end{equation}
\ignore{
If $g : G \rightarrow \Complex$, then the convolution of $f$ and $g$ is
\begin{equation}
(f \star g)(y) = \sum_{x \in G} f(yx^{-1})g(x).
\end{equation}
The Fourier transform of a convolution is the product of the Fourier transforms:
\begin{equation}
\widehat{f \star g}(\rho) = \hat{f}(\rho)\hat{g}(\rho).
\end{equation}
}
Fourier inversion reconstructs $f$ from its Fourier transform at all
irreducible representations $\rho_{1},\ldots,\rho_{h}$ of $G$ with dimensions $d_{1},\ldots,d_{h}$, respectively:
\begin{equation}
f(x) = \frac{1}{|G|} \sum_{j=1}^{h} d_{j} Tr(\rho_{j}(x^{-1})\hat{f}(\rho_{j})).
\end{equation}
For each irreducible representation $\rho_{j}$, 
let $D_{j} = diag(\hat{f}(\rho_{j}))$ be a $d_{j}^{2} \times d_{j}^{2}$ block matrix.
Now let $D = diag(D_{1},\ldots,D_{h})$. 
Define the vector $\psi_{j}$ of length $d_{j}^{2}$ as
$\psi_{j}(x) = (\sqrt{d_{j}}/|G|)(\rho_{j}(x)[s,t])_{1 \le s,t \le d_{j}}$
and the vector $\psi(x) = (\psi_{j}(x))_{j=1}^{h}$ of length $|G|$. 

If $f: G \rightarrow \Complex$ is a class function over the group $G = \{x_{1},\ldots,x_{N}\}$, i.e.,
$f$ is constant on the conjugacy classes of $G$,
then, $M_{G}^{f}$ is {\em unitarily diagonalized} by $\Psi = (\psi(x_{1}), \ldots, \psi(x_{N}))$, i.e.,
\begin{equation}
M_{G}^{f} = \Psi^{\dagger} D \Psi,
\end{equation}
where, for $j \in [n]$, $D_{j} = \lambda_{j} I_{d_{j}^{2}}$, 
$\chi_{j}(x) = Tr(\rho_{j}(x))$ is the character of $\rho_{j}$ at $x$,
and the eigenvalue is 
\begin{equation} \label{eqn:group_lambda}
\lambda_{j} = \frac{1}{d_{j}}\sum_{x \in G} f(x)\overline{\chi_{j}}(x).
\end{equation}

\par\noindent 
From the above, if a group $G$ contains a representation of dimension greater than $1$,
then the matrix $M^{f}_{G}$, for some class function $f$, has non-distinct eigenvalues. For Abelian groups,
where all representations are of dimension $1$, all eigenvalues of $M_{G}^{f}$ are not guaranteed to be distinct,
as we observe next.

\subsection{Abelian circulants and average uniform mixing}

Given the Hadamard matrix $H_{2} = \begin{pmatrix} 1 & 1 \\ 1 & -1 \end{pmatrix}$ of order $2$,
let $H_{n} = \begin{pmatrix} H_{n-1} & H_{n-1} \\ H_{n-1} & -H_{n-1} \end{pmatrix}$ be the Hadamard
matrix of order $n > 2$ defined recursively in terms the Hadamard matrix of order $n-1$.
We call graph $G$ a Hadamard circulant if it is diagonalized by some Hadamard matrix $H_{n}$.
Alternatively, these are $G$-circulant matrices for $G = \Int_{2} \times \ldots \times \Int_{2}$.
The following result was a main result in \cite{mr}.

\begin{theorem} \label{thm:moore-russell} (Moore, Russell \cite{mr}) \\
The continuous-time quantum walk on the $n$-cube is instantaneous exactly uniform for times
$t = k \frac{\pi}{4} n$, for odd positive integers $k$. Also, there is $\epsilon > 0$ such that
no $\epsilon$-average mixing exists.
\end{theorem}

\par\noindent Equation (\ref{eqn:aakv}) of Aharonov \etal suggests that a graph with distinct eigenvalues 
diagonalized by Hadamard matrices has average uniform mixing; but these graphs have spectral gap zero.

\begin{lemma} \label{lemma:hadamard}
Let $G$ be a graph diagonalized by 
$H_{n}$, for $n > 2$.
Then $G$ has spectral gap zero.
\end{lemma}
\prf 
Consider the characters of $\Int_{2}^{n}$ defined for each $a \in \Int_{2}^{n}$ as 
$\chi_{a}(x) = \prod_{j=1}^{n} (1-2a_{j}x_{j})$. From Equation (\ref{eqn:group_lambda}), we get
$\lambda_{a} = \sum_{x \in \Int_{2}^{n}} f(x)\chi_{a}(x)$,
where $f: \Int_{2}^{n} \rightarrow \zo$ defines the first column of the adjacency matrix of $G$. 
Let $|f| = \{x \neq 0_{n} : f(x) = 1\}$. Assume that $|f| < 2^{n}-1$, otherwise we get the complete graph
which has only $2$ distinct eigenvalues.
If $|f|$ is even, then $\lambda_{a} \in \{0, \pm 2, \ldots, \pm |f|\}$. Since the eigenvalues can take at most 
$|f|+1 < 2^{n}$ values, by the pigeonhole principle, there exist two non-distinct eigenvalues.
If $|f|$ is odd, then $\lambda_{a} \in \{\pm 1, \pm 3, \ldots, \pm |f|\}$. Similarly, the eigenvalues range
on at most $|f| < 2^{n}-1$ values, and again there exist two non-distinct eigenvalues.
\qed

\begin{theorem} \label{thm:abelian_circulant}
For Abelian $G$, no $G$-circulant, except for $K_{2}$, is average uniform mixing.
\end{theorem}
\prf Let $G = \Int_{n_{1}} \times \ldots \times \Int_{n_{k}}$ be an Abelian group.
If all elements of $G$ have order $2$ (except for the identity), we appeal to Lemma \ref{lemma:hadamard}.
Otherwise, fix $a \in G$ with order greater than $2$.
The character corresponding to $a$ is $\chi_{a}(x) = \prod_{j=1}^{k} \chi_{a_{j}}(x_{j})$. 
From Equation (\ref{eqn:group_lambda}),
\begin{equation}
\lambda_{a} 
	= \sum_{x \neq 0} f(x)\overline{\chi}_{a}(x)
	= \sum_{x \neq 0} f(x)\overline{\chi}_{-a}(-x)
	= \sum_{x \neq 0} f(-x)\overline{\chi}_{-a}(-x)
	= \lambda_{-a}.
\end{equation}
Thus, the spectral gap of $M_{G}^{f}$ is zero.
Finally, since $G$ is Abelian, its characters are complex roots of unity;
thus, applying Lemma \ref{lemma:aakv}, we obtain the claim.
\qed\\

\par\noindent The above theorem implies that the $n$-cube and the standard circulant graphs
are not average uniform mixing, as summarized in the following corollaries. Recall that
Moore and Russell \cite{mr} proved a stronger non-uniform mixing on the $n$-cube.

\begin{corollary}
No $\Int_{2}^{n}$-circulant and no $\Int_{n}$-circulant, except for $K_{2}$, is average uniform mixing.
\end{corollary}

\ignore{
\begin{figure}[h]
\[\begin{xy} /r10mm/:
 ,{\xypolygon4"A"{~={0}\dir{*}}}
 ,+/r40mm/
 ,{\xypolygon4"M"{~={0}\dir{*}}}
 ,+(0.0,1.0), ,{\xypolygon4"N"{~={0}\dir{*}}}
 ,"M1";"N1"**@{-},"M2";"N2"**@{-},"M3";"N3"**@{-},"M4";"N4"**@{-}
 ,+/r35mm/
 ,{\xypolygon4"F"{\dir{*}}}
 ,+(.8,1.0), {\xypolygon4"B"{\dir{*}}}
 ,"F1";"B1"**@{-},"F2";"B2"**@{-},"F3";"B3"**@{-},"F4";"B4"**@{-}
 ,+/r30mm/
\end{xy}\]
\caption{Two group-theoretic generalizations of the $n$-cube.
From left to right: (a) the $2$-cube, $\Int_{2} \times \Int_{2}$ or $\Int_{2} \rtimes \Int_{2}$; 
(b) the $(\Int_{2} \times \Int_{2} \times \Int_{2})$-circulant; (c) the $\Int_{4} \rtimes \Int_{2}$-bunkbed.}
\label{figure:group-cube}
\end{figure}
}

\subsection{$\Int_{n}$-circulants and average almost uniform mixing}

Next we consider average almost uniform mixing of circulants. We observe that the complete cycle and
the complete graphs form opposite extremes in behavior with respect to average almost uniform mixing.

\begin{theorem}
The complete cycle $C_{n}$ is average $(1/n)$-uniform mixing.
\end{theorem}
\prf
Let $\omega = \exp(2\pi i/n)$. Using Equation (\ref{eqn:aakv}), we have
\begin{equation}
\overline{P}(\ell) 
	= \frac{1}{n^2}\sum_{j,k=0}^{n-1} \omega^{(j-k)\ell}\iverson{\lambda_{j} = \lambda_{k}}
	= \frac{1}{n} + \frac{1}{n^{2}}\sum_{j \neq k} \omega^{(j-k)\ell}\iverson{\lambda_{j} = \lambda_{k}}.
\end{equation}
A result of Diaconis and Shahshahani (see \cite{diaconis-book}; also \cite{mr}) states 
$||\overline{P} - U|| \le \frac{1}{4}\sum_{\rho} |\widehat{\overline{P}}(\rho)|^{2}$,
where the sum is over nontrivial irreducible representations.
The characters of $\Int_{n}$ are given by $\chi_{a}(x) = \omega^{ax}$, and thus, for $a \neq 0$,
\begin{equation}
\widehat{\overline{P}}(a) = \sum_{\ell} \overline{P}(\ell)\chi_{a}(\ell) 
	= \frac{1}{n^{2}}\sum_{j \neq k: \ \lambda_{j} = \lambda_{k}} \sum_{\ell} \omega^{(j-k+a)\ell} 
	= \frac{1}{n}.
\end{equation}
The last equality is because there is a unique pair $(j,k)$ such that $j-k+a=0$; this pair contributes $n$ to the sum while
the other pairs contribute $0$ to the sum. Therefore, 
$||\overline{P} - U|| \le (n-1)/4n^{2} < 1/4n$.
\qed

\begin{theorem} (Ahmadi \etal \cite{abtw})
The complete graph $K_{n}$ is not average $(1/n^{O(1)})$-uniform mixing.
\end{theorem}
\prf
As shown in \cite{abtw}, for $\ell \neq 0$, we have $\overline{P}(\ell) = 2/n^{2}$, 
and $\overline{P}(0) = 1-2(n-1)/n^{2}$. Thus $||\overline{P}-U|| = 2(1-1/n)(1-2/n) \sim 2$.
\qed\\

\par\noindent 
Borrowing a terminology from spectral graph theory, a graph $G$ is called {\em type $k$} if it has
$k$ distinct eigenvalues. The complete graph $K_{n}$ is type $2$, whereas the complete cycle $C_{n}$
is type $1 + \lfloor n/2 \rfloor$. 
Consider a random 
$\Int_{n}$-circulant $C(n,1/2)$, where we set
$f(0) = 0$, 
and for $j \in \{1,\ldots,\lfloor n/2 \rfloor\}$, we choose $f(j) = f(n-j)$ 
to be independent Bernoulli random $\zo$-variables. 
The {\em expected} eigenvalues of $C(n,1/2)$ satisfy
$\ep[\lambda_{0}] = \lfloor n/2 \rfloor$, 
and $\ep[\lambda_{j}] = -\frac{1}{2}$, for $j \neq 0$.
By standard concentration bounds, the values $\lambda_{j}$, for $j \neq 0$, 
are concentrated around its mean (and median). So, the average type of a random circulant is near $K_{n}$.
A detailed analysis on the {\em type spectra} of random $\Int_{n}$-circulants would be helpful.

\begin{conjecture}
Almost all Abelian circulants are not average $(1/n^{O(1)})$-uniform mixing.
\end{conjecture}

\section{Group-theoretic bunkbeds} 

Given a group $G$ and a set $S \subseteq G$, a Cayley graph $\Gamma(G,S)$ is defined on the vertex set $V = G$
and edge set $E = \{(g,h) : \exists s \in S \ h = gs\}$. The graph $\Gamma(G,S)$ is undirected if $S$ is closed 
under taking inverses, i.e., for all $s \in S$, $s^{-1} \in S$, 
and is connected if $S$ is a generating set, i.e., $G = \lip S \rip$. 
Some additional requirements might be placed on $S$, for example, $S$ must be a {\em minimal} generating set, i.e.,
for any $s \in S$, $\lip S \setminus \{s\}\rip \neq G$, or $S$ must be a conjugacy class (see \cite{gw}).

\begin{definition} (Bunkbed graph) \\
Let $G$ be a group that is a semidirect product $A \rtimes \Int_{2}$, 
where $A = \lip g_{j}  : j \in [k]\rip$ is a group generated by $k$ generators $\{g_{1},\ldots,g_{k}\}$.
Suppose the relations $a^2 = e$, $a g_{j} a^{-1} \in \{g_{j}^{-1}, g_{j}\}$, hold for each $j \in [k]$.
If $S = \{a\} \cup \{g_{j}, g_{j}^{-1} \ : \ j \in [k]\}$, then 
the Cayley graph $\Gamma(G,S)$ is called a {\em (group-theoretic) bunkbed} of $G$.
\end{definition}

The vertex set of a bunkbed $G = A \rtimes \Int_{2}$ is $V_{G} = \zo \times V_{A}$, 
where $V_{A}$ is the vertex set of the Cayley graph $G_{A}$ of $A$. 
The bunkbed $G$ consists of two isomorphic copies of $G_{A}$ connected by parallel edges. 
Let $P_{t}(b,\ell)$ be the probability of observing the particle at vertex $(b,\ell)$ at time $t$, 
where $b \in \{0,1\}$ and $\ell \in V_{A}$; the corresponding average probability is denoted $\overline{P}(b,\ell)$.
We consider also the conditional average probability $\overline{P}_{b}(\ell)$ over the two partitions, for each $b \in \zo$.

\begin{theorem} \label{thm:average-bunkbed}
In a continuous-time quantum walk on the bunkbed $G = A \rtimes \Int_{2}$, 
we have $\overline{P}_{0} \equiv \overline{P}_{1}$.
\end{theorem}
\prf
Let $A_{n}$ be the adjacency matrix of the 
$\Gamma(A,\lip g_{j} : j \in [k]\rip)$ with $n$ vertices.
The adjacency matrix of $G$ is $H = I_{2} \otimes A_{n} + X_{2} \otimes I_{n}$,
where $I_{k}$ is the $k \times k$ identity matrix and $X_{2}$ is the Pauli $X$ matrix. 
If $\ket{z_{0}}$ and $\ket{z_{1}}$ are the common eigenvectors of $I_{2}$ and $X_{2}$
(both being circulants), and $\ket{\alpha_{j}}$ are the eigenvectors of $A_{n}$
with eigenvalues $\lambda_{j}$, $j \in [n]$, then, assuming the continuous-time quantum walk starts at
vertex $(0,0_{n})$, 
\begin{eqnarray*}
\ket{\psi(t)} 
	& = & e^{-iHt}\ket{0}\ket{0_{n}} 
		= e^{-iHt} \sum_{b=0}^{1}\sum_{j=0}^{n} \braket{z_{b}}{0}\braket{\alpha_{j}}{0_{n}} \ket{z_{b}}\ket{\alpha_{j}} \\
	& = & \sum_{b=0}^{1}\sum_{j=0}^{n} \braket{z_{b}}{0}\braket{\alpha_{j}}{0_{n}} 
		e^{-it(I_{2} \otimes A_{n})} e^{-it(X_{2} \otimes I_{n})} \ket{z_{b}}\ket{\alpha_{j}},
		\ \ \mbox{\em since $I_{2} \otimes A_{n}$ and $X_{2} \otimes I_{n}$ commute} \\
	& = & e^{-it}\sum_{j=0}^{n} \braket{z_{0}}{0}\braket{\alpha_{j}}{0_{n}} 
		e^{-it \lambda_{j}} \ket{z_{0}}\ket{\alpha_{j}} 
		+ 
		e^{it}\sum_{j=0}^{n} \braket{z_{1}}{0}\braket{\alpha_{j}}{0_{n}} 
		e^{-it \lambda_{j}} \ket{z_{1}}\ket{\alpha_{j}} \\
\end{eqnarray*} 
Since $\braket{z_{0}}{0} = \braket{z_{1}}{0} = 1/\sqrt{2}$, we have
\begin{eqnarray*}
\ket{\psi(t)}
	& = & \frac{1}{\sqrt{2}} \left( e^{-it}\ket{z_{0}} + e^{it}\ket{z_{1}} \right) 
		\otimes \sum_{j=0}^{n} \braket{\alpha_{j}}{0_{n}} e^{-it \lambda_{j}} \ket{\alpha_{j}} \\
	& = & (\cos(t) \ket{0} - i\sin(t) \ket{1})
		\otimes \sum_{j=0}^{n} \braket{\alpha_{j}}{0_{n}} e^{-it \lambda_{j}} \ket{\alpha_{j}}.
\end{eqnarray*}
For $b \in \zo$ and $\ell \in [n]$, the probability of observing the particle at vertex $(b,\ell)$ at time $t$ is
\begin{equation}
P_{t}(b,\ell) = [(1-b)\cos^{2}(t) + b\sin^{2}(t)] \sum_{j,k=0}^{n-1} e^{-it(\lambda_{j}-\lambda_{k})} 
	\braket{\alpha_{j}}{0_{n}}\braket{0_{n}}{\alpha_{k}} \braket{\ell}{\alpha_{j}}\braket{\alpha_{k}}{\ell}.
\end{equation}
\ignore{
\begin{equation}
P_{t}(1,\ell) = \sin^{2}(t) \sum_{j,k=0}^{n-1} e^{-it(\lambda_{j}-\lambda_{k})} 
	\braket{\alpha_{j}}{0_{n}}\braket{0_{n}}{\alpha_{k}} \braket{\ell}{\alpha_{j}}\braket{\alpha_{k}}{\ell}.
\end{equation}
}
Given that 
$\lim_{T \rightarrow \infty} \frac{1}{T}\int_{0}^{T} e^{-it \Delta} = \iverson{\Delta = 0}$ and 
$\lim_{T \rightarrow \infty} \frac{1}{T}\int_{0}^{T} \cos^{2}(t) \dt = 
\lim_{T \rightarrow \infty} \frac{1}{T}\int_{0}^{T} \sin^{2}(t) \dt = \frac{1}{2}$, 
for $\ell \in [n]$,
\begin{equation}
\overline{P}(0,\ell) = \overline{P}(1,\ell) 
	= \frac{1}{2}\sum_{j,k=0}^{n} \braket{\alpha_{j}}{0_n}\braket{0_n}{\alpha_{k}}
	\braket{\ell}{\alpha_{j}}\braket{\alpha_{k}}{\ell}
	\iverson{\lambda_{j} = \lambda_{k}}.
\end{equation}
\qed

\par\noindent We conjecture that no Abelian bunkbed, except for $K_{2}$, is average uniform mixing.

\begin{conjecture} 
For Abelian $G$, no $G$-bunkbed, i.e., $G \rtimes \Int_{2}$, except for $K_{2}$, is average uniform mixing.
\end{conjecture}

\par\noindent
A graph $G = A \rtimes \Int_{2}$ is called a {\em circulant bunkbed} if $A$ is a circulant graph. 
The only known circulant bunkbeds with instantaneous uniform mixing are the $n$-cube and 
$K_{4} \rtimes \Int_{2} \rtimes \ldots \rtimes \Int_{2}$ 
(given that $K_{4}$ also mixes uniformly at multiples of $\pi/4$; see \cite{abtw}).
It seems plausible that these are the only ones.
\ignore{
We conjecture that these are the only ones.

\begin{conjecture}
No circulant bunkbed, 
except for $\Int_{2} \times \ldots \times \Int_{2}$ and $K_{4} \rtimes \Int_{2} \rtimes \ldots \rtimes \Int_{2}$, 
is instantaneous uniform mixing.
\end{conjecture}
}

\ignore{
\begin{figure}[t]
\[\begin{xy} /r10mm/:
 ,{\xypolygon2"I"{~={0}\dir{*}}}
 ,+/r30mm/
 ,+(1.0,0.5), {\xypolygon4"P"{\dir{*}}}
 ,+(0.75,0.25), {\xypolygon4"Q"{\dir{*}}}
 ,"P1";"Q1"**@{-},"P2";"Q2"**@{-},"P3";"Q3"**@{-},"P4";"Q4"**@{-}
 ,+/r20mm/
 ,+(1.0,0.5), {\xypolygon4"F"{\dir{*}}}
 ,+(0.75,0.25), {\xypolygon4"B"{\dir{*}}}
 ,"F1";"B1"**@{-},"F2";"B2"**@{-},"F3";"B3"**@{-},"F4";"B4"**@{-}
 ,"F1";"F3"**@{-},"F2";"F4"**@{-},"B1";"B3"**@{-},"B2";"B4"**@{-}
\end{xy}\]
\caption{Some circulant bunkbeds:
(a) $K_{2}$;
(b) $\Int_{4} \rtimes \Int_{2}$-bunkbed;
(c) $K_{4} \rtimes \Int_{2}$-bunkbed.}
\label{figure:circulant-bunkbed}
\end{figure}
}

\subsection{Dihedral groups}

Let $D_{n} = \Int_{n} \rtimes \Int_{2}$ be the dihedral group of order $2n$ defined by two generators $a$ and $b$ where
$D_{n} = \lip a,b : a^{n} = b^{2} = 1, \ bab = a^{-1}\rip$.
The Cayley graph $G = \Gamma(D_{n},S = \{b,a,a^{-1}\}$ 
is a bunkbed graph with two isomorphic cycles $C_{n}$ joined by parallel edges.
If $S = \{b,ab\}$, then $G$ is the complete cycle $C_{2n}$ of length $2n$. 
Both choices of $S$ are minimal but neither are conjugacy classes. 
Some non-minimal choices of $S$ include the trivial $S = D_{n}$ and the conjugacy class $S = \{b,ab,a^{2}b,\ldots,a^{n-1}b\}$ 
(only for $n$ even) which yield the complete graph $K_{2n}$ and the complete bipartite graph $K_{n,n}$, respectively.
None of these produce graphs with average uniform mixing for the continuous-time quantum walk.

On Abelian groups $G$, a class function $f$ may assign arbitrary values to elements of $G$.
On groups of the form $G \rtimes \Int_{2}$, we ask if a Boolean class function could induce a bunkbed graph
through $M_{G \rtimes \Int_{2}}^{f}$. 
A negative answer would show a limitation of the group circulant method in analyzing bunkbeds.


\section{Average mixing on paths}

We consider a rare natural class of graphs with distinct eigenvalues --
the complete paths $P_{n}$ of order $n \ge 2$.
A classical discrete lazy random walk on $P_{n}$ has a stationary distribution $\pi$ defined
as $\pi(0) = \pi(n-1) = 1/2(n-1)$, and $\pi(j) = 1/(n-1)$, for $1 \le j \le n-2$.

\begin{theorem}
No complete path, except for $K_{2}$, is average classical mixing.
\end{theorem}
\prf
The complete spectrum of $P_{n}$ is given by Spitzer \cite{spitzer}.
For $j \in [n]$, the eigenvalue $\lambda_j$ of $P_{n}$ and its eigenvector $\ket{v_j}$ are given by
\begin{equation}
\lambda_j = \cos\left(\frac{j+1}{n+1}\pi\right), \ \ \
\braket{\ell}{v_j} = \sqrt{\frac{2}{n+1}}\sin\left(\frac{j+1}{n+1}\pi(\ell + 1)\right).
\end{equation}
Note 
\ignore{
\begin{equation}
\ket{\psi(t)} 
	= e^{-itP_n}\ket{0} 
	= \sqrt{\frac{2}{n+1}} \sum_j \sin\left(\frac{j+1}{n+1}\pi\right) e^{-it\lambda_j}\ket{v_j},
\end{equation}
and
\begin{equation}
\bra{0}e^{-itP_n}\ket{0} = \frac{2}{n+1} \sum_j \sin^2\left(\frac{j+1}{n+1}\pi\right) e^{-it\lambda_j}.
\end{equation}
Thus
}
\begin{equation}
P_{t}(0) = |\bra{0}e^{-itP_n}\ket{0}|^2 
	= \frac{4}{(n+1)^2} 
		\sum_{j,k} \sin^2\left(\frac{j+1}{n+1}\pi\right) \sin^2\left(\frac{k+1}{n+1}\pi\right) 
		e^{-it(\lambda_j-\lambda_k)}.
\end{equation}
Since all eigenvalues of $P_n$ are distinct, the average probability of the starting vertex $0$ is
\begin{eqnarray*}
\overline{P}(0) 
	& = & \lim_{T \rightarrow \infty} \frac{1}{T}\int_{0}^{T} |\bra{0}e^{-itP_n}\ket{0}|^2 \\
	& = & \frac{4}{(n+1)^2}
		\sum_{j,k} \sin^2\left(\frac{j+1}{n+1}\pi\right) \sin^2\left(\frac{k+1}{n+1}\pi\right) 
		\lim_{T \rightarrow \infty} \frac{1}{T}\int_{0}^{T} e^{-it(\lambda_j-\lambda_k)} \\
	& = & \frac{4}{(n+1)^2} \sum_{j} \sin^4\left(\frac{j+1}{n+1}\pi\right) \\
	& \leq & \frac{4}{(n+1)^2} \ \ \frac{3\pi}{8}, 
\end{eqnarray*}
since $\int \sin^4(x) \mathsf{d}x = -\frac{1}{4}\sin^3(t)\cos(t) + \frac{3t}{8} - \frac{3}{16}\sin(2t) + C$.
This implies that $\overline{P}(0) < \frac{1}{2(n-1)}$, for $n > 5$.
Given that $\overline{P}(0) = 1/2(n-1)$, if $n=2$, and $\overline{P}(0) > 1/2(n-1)$, if $n=3,4,5$,
we have the claim.
\qed\\

\par\noindent
We were unable to determine yet if $P_{n}$ is average $\epsilon$-classical mixing.

\section*{Acknowledgments}

\noindent
This work is supported in part by NSF DMR-0121146 and (REU) DMS-0097113. 
We thank P. H\o{}yer for helpful discussions in the early stages of this research.
Part of this work was done while C. T. visited the University of Calgary;
he is grateful for the kind hospitality of R. Cleve, P. H\o{}yer, and J. Watrous.

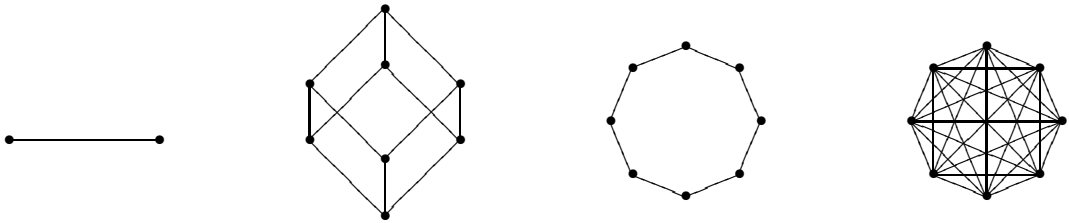
\begin{figure}[h]
\[\begin{xy} /r10mm/:
 ,{\xypolygon2"S"{~={0}\dir{*}}}
 ,+/r40mm/
 ,{\xypolygon4"M"{~={0}\dir{*}}}
 ,+(0.0,0.75),{\xypolygon4"N"{~={0}\dir{*}}}
 ,"M1";"N1"**@{-},"M2";"N2"**@{-},"M3";"N3"**@{-},"M4";"N4"**@{-}
 ,+/r40mm/
 ,+(0.0,0.5),{\xypolygon8"C"{~={0}\dir{*}}}
 ,+/r40mm/
 ,{\xypolygon8"K"{~={0}\dir{*}}}
 ,"K1";"K3"**@{-},"K2";"K4"**@{-},"K3";"K5"**@{-},"K4";"K6"**@{-},"K5";"K7"**@{-},"K6";"K8"**@{-},"K7";"K1"**@{-},"K8";"K2"**@{-}
 ,"K1";"K4"**@{-},"K2";"K5"**@{-},"K3";"K6"**@{-},"K4";"K7"**@{-},"K5";"K8"**@{-},"K6";"K1"**@{-},"K7";"K2"**@{-},"K8";"K3"**@{-}
 ,"K1";"K5"**@{-},"K2";"K6"**@{-},"K3";"K7"**@{-},"K4";"K8"**@{-}
\end{xy}\]
\caption{Some Abelian circulants and their quantum mixing characteristics.
From left to right: 
(a) the smallest circulant $K_{2}$; 
{\em instantaneous and average uniform}.
(b) the $(\Int_{2} \times \Int_{2} \times \Int_{2})$-circulant, i.e., $3$-cube; 
{\em instantaneous but not average uniform}.
(c) the cycle $\Int_{8}$-circulant, i.e., $C_{8}$; 
unknown instantaneous behavior, but 
{\em average $(1/n^{O(1)})$-uniform}.
(d) the complete $\Int_{8}$-circulant, i.e., $K_{8}$; 
{\em neither instantaneous nor average $(1/n^{O(1)})$-uniform}.
}
\label{figure:abelian-circulant}
\end{figure}


\begin{thebibliography}{000}

\bibitem{aakv}
D. Aharonov, A. Ambainis, J. Kempe, and U. Vazirani (2001),
{\it Quantum Walks on Graphs},
in Proc. 33rd ACM Ann. Symp. Theory of Computing (ACM Press), 50-59.

\bibitem{abtw}
A. Ahmadi, R. Belk, C. Tamon, C. Wendler (2002),
{\it Mixing in continuous quantum walk on graphs}, quant-ph/0209106.

\bibitem{alw}
N. Alon, A. Lubotzky, A. Wigderson (2001),
{\it Semi-direct product in groups and Zig-zag product in graphs: Connections and applications},
in Proc. 42nd IEEE Ann. Symp. Foundations of Computer Science (IEEE Society), 630-637.

\bibitem{abnvw}
A. Ambainis, E. Bach, A. Nayak, A. Viswanath, and J. Watrous (2001),
{\it One-dimensional Quantum Walks},
in Proc. 33rd ACM Ann. Symp. Theory of Computing (ACM Press), 60-69.

\bibitem{ccdfgs}
A. Childs, E. Deotto, R. Cleve, E. Farhi, S. Gutmann, D. Spielman (2003),
{\it Exponential algorithmic speedup by quantum walk},
in Proc. 35th ACM Ann. Symp. Theory of Computing (ACM Press), 59-68.

\bibitem{cfg}
A. Childs, E. Farhi, and S. Gutmann (2002),
{\it An example of the difference between quantum and classical random walks},
Quantum Information Processing 1, 35.

\bibitem{diaconis-book}
P. Diaconis (1988),
{\it Group Representations in Probability and Statistics},
Institute of Mathematical Statistics.

\bibitem{diaconis}
P. Diaconis (1990),
{\it Patterned Matrices},
in Matrix Theory and Applications, Proceedings of Symposia in Applied Mathematics, Volume 40,
American Mathematical Society, 37-58.

\bibitem{fg}
E. Farhi and S. Gutmann (1998),
{\it Quantum computation and decision trees},
Phys. Rev. A 58. 

\bibitem{gw}
H. Gerhardt and J. Watrous (2003),
{\it Continuous-time quantum walks on the symmetric group},
in quant-ph/0305182. To appear in RANDOM'03.

\bibitem{haggstrom}
O. H\"{a}ggstrom (2003),
{\it Probability on bunkbed graphs},
in Formal Power Series and Algebraic Combinatorics (FPSAC 2003), K. Eriksson and S. Linusson, eds.
(Link\"{o}pings universiteit), 19-27.

\bibitem{k}
J. Kempe (2003),
{\it Quantum Random Walks Hit Exponentially Faster}, in quant-ph/0205083. To appear in RANDOM'03.

\bibitem{mr}
C. Moore and A. Russell (2002),
{\it Quantum Walks on the Hypercube}, in Proc. 6th Int. Workshop 
Randomization and Approximation in Computer Science (RANDOM'02). Also in quant-ph/0104137.

\bibitem{serre}
J.-P. Serre (1977),
{\it Linear Representations of Finite Groups}, Springer.

\bibitem{spitzer}
F. Spitzer (1976),
{\em Principles of Random Walk}, 2nd ed., Springer.

\end{thebibliography}
\end{document}